\documentclass[10pt,a4paper]{article}

\usepackage[latin1]{inputenc}
\usepackage{fourier}
\usepackage{graphicx}
\usepackage{amsmath}
\usepackage{amsfonts}
\usepackage[latin1]{inputenc}
\usepackage[T1]{fontenc}
\usepackage{verbatim}
\usepackage{latexsym}
\usepackage{natbib}
\usepackage{bm}
\usepackage[labelfont=bf]{caption}
\usepackage{multirow}
\usepackage{array}
\usepackage{enumitem}

\usepackage{hyperref}
\hypersetup{backref,colorlinks=true,citecolor=blue}

\usepackage{alltt}
\usepackage{color}                       % For color text: \color command
\definecolor{grey}{rgb}{0.9,0.9,0.9}

\newcommand{\mdeg}[1]{^\circ#1}

\usepackage{etoolbox}
\makeatletter
\patchcmd{\NAT@citex}
  {\@citea\NAT@hyper@{%
     \NAT@nmfmt{\NAT@nm}%
     \hyper@natlinkbreak{\NAT@aysep\NAT@spacechar}{\@citeb\@extra@b@citeb}%
     \NAT@date}}
  {\@citea\NAT@nmfmt{\NAT@nm}%
   \NAT@aysep\NAT@spacechar\NAT@hyper@{\NAT@date}}{}{}
\patchcmd{\NAT@citex}
  {\@citea\NAT@hyper@{%
     \NAT@nmfmt{\NAT@nm}%
     \hyper@natlinkbreak{\NAT@spacechar\NAT@@open\if*#1*\else#1\NAT@spacechar\fi}%
       {\@citeb\@extra@b@citeb}%
     \NAT@date}}
  {\@citea\NAT@nmfmt{\NAT@nm}%
   \NAT@spacechar\NAT@@open\if*#1*\else#1\NAT@spacechar\fi\NAT@hyper@{\NAT@date}}
  {}{}
\makeatother

\begin{document}

\title{Polar Field Correction for HMI Line-of-Sight Synoptic Data}

\author{Xudong Sun$^{1,2}$ for the HMI Team \\
\small $^{1}$ HEPL, Stanford University, Stanford, CA 94305 (\href{mailto:xudong@sun.stanford.edu}{xudong@sun.stanford.edu})\\
\small $^{2}$ Now at: IfA, University of Hawaii at Manoa, Pukalani, HI 96768}

%\date{\normalsize \textit{\today}}
\date{\vspace{-3ex}}

\maketitle

\begin{abstract}
This document provides some technical notes on the polar field correction scheme for the HMI synoptic maps and daily updated synchronic frames. It is intended as a reference for the new data products and for some minor updates on our previous scheme for MDI \citep{sun2011}.
\end{abstract}

\section{Correction Scheme}
\label{sec:synop}

\subsection{HMI Synoptic Data}

\begin{figure}[b!]
\centerline{\includegraphics[scale=0.9]{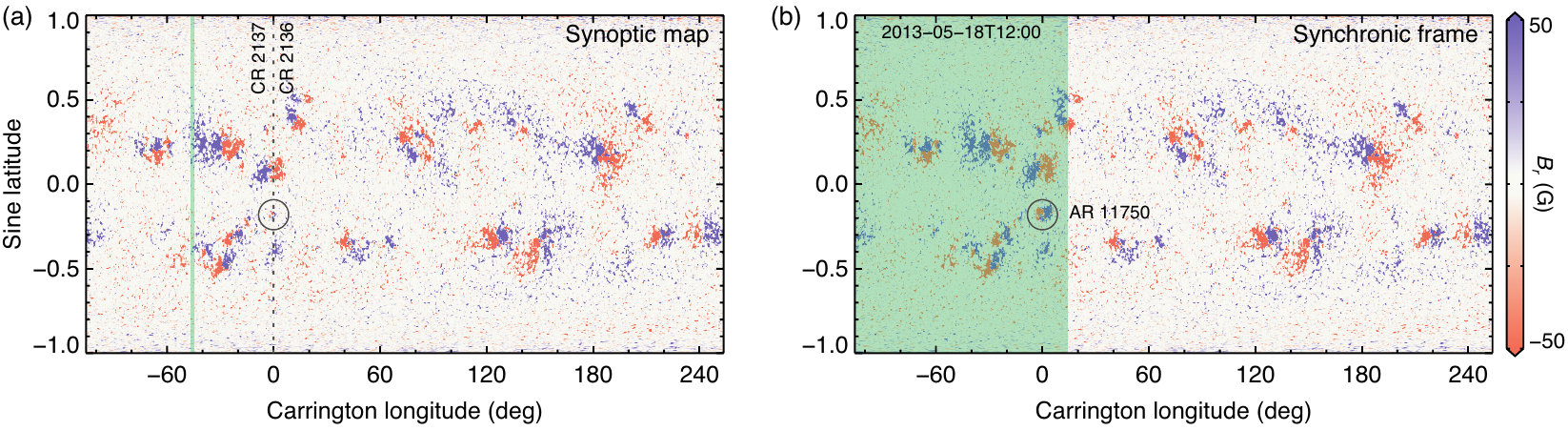}}
\caption{\textbf{Example of HMI synoptic data.} (a) Full-CR synoptic maps assembled from CR 2136 and 2137. (b) Daily-updated synchronic frame for 2013-05-28T12:00. The green-shaded bands show the region where observations from 2013-05-28T12:00 are used: $2.2\mdeg$ in the synoptic map and $120\mdeg$ in the synchronic frame. The circle shows NOAA AR 11750. It started emerging on the western hemisphere and thus is not present in the synoptic map.\label{f:illus}}
\end{figure}

The full Carrington-rotation (CR) HMI synoptic map has dimensions of $3600\times1440$, equally spaced in longitude ($\phi$) and sine-latitude (Figure~\ref{f:illus}(a)). The map is in a cylindrical equal area (CEA) projection, where each pixel represents the same area on the Sun \citep{thompson2006}. Near the equator the pixel size is about $0.1\mdeg\times0.08\mdeg$. Each column of the map represents a particular Carrington longitude at its central meridian passage, averaged from several magnetograms. The averaging window size is 20 frames, typically 4 hrs of 720 s cadence magnetograms, covering $2.2\mdeg$ longitude.

The \textbf{synoptic maps} are constructed as follows. Each line-of-sight (LoS) magnetogram ($B_l$) is first converted to a radial field ($B_r$) map using the assumption that the photospheric field is purely radial. These radial values are spatially interpolated onto a higher resolution Carrington coordinate map. The radial field values in the Carrington system are averaged to compute the radial synoptic map, and finally the LoS synoptic map is computed from the radial synoptic map. The noise level is $2.3\mathrm{~G}$ for the $B_l$ maps \citep[\href{http://jsoc.stanford.edu/ajax/lookdata.html?ds=hmi.synoptic_ml_720s}{\texttt{hmi.synoptic\_ml\_720s}};][]{liuy2012}. It scales approximately as the inverse of cosine latitude in the $B_r$ maps (\href{http://jsoc.stanford.edu/ajax/lookdata.html?ds=hmi.synoptic_mr_720s}{\texttt{hmi.synoptic\_mr\_720s}}).

The solar rotation axis has a $7.25\mdeg$ tilt angle, so the polar region is partly unobserved. We also discard data outside 0.998 apparent solar disk radii due to high noise. These two factors lead to a maximum of 15 rows missing from the synoptic maps, down to $\pm78\mdeg$ latitude. The maximum tilt toward the Earth occurs in early September and March for the north and the south, respectively.

%%%%%%%%%%%

%%%%%%%%%%%

The daily-updated \textbf{synchronic frame} (\href{http://jsoc.stanford.edu/ajax/lookdata.html?ds=hmi.mrdailysynframe_720s}{\texttt{hmi.mrdailysynframe\_720s}}) is designed to provide a full-Sun snapshot that better represents the front side conditions (Figure~\ref{f:illus}(b)). It updates a synoptic map with new observations centered at time $T_C$, or Carrington longitude $\phi_C$. To do so, we rotate the existing full-CR maps such that the column with Carrington longitude $\phi_C$ (of previous CR) is $60\mdeg$ west of the leading (left) edge. We then average 20 new magnetograms whose central meridians are closest to $\phi_C$, and use the averaged data within a $\pm60\mdeg$ longitudinal window to replace the corresponding part in the synoptic map ($120\mdeg$ west from the leading edge). Near the boundary of the updated part, more data may be missing at high latitudes (to a maximum of 31 rows, or $\pm73\mdeg$ latitude). One synchronic frame is produced each day at about 12:00 UTC.

For a synoptic map specified by CR number $N$ and a synchronic frame specified by $T_C$ (corresponding to Carrington longitude $\phi_C$ of CR number $N$), we have the keywords for the coordinate system listed in Table~\ref{tbl:eph}.

%%%%%%%%%%%

\begin{table}[t]
\begin{center}
\caption{Ephemeris keywords for HMI synoptic data \label{tbl:eph}}
\begin{tabular}{c|cc|cc|cc|cc}
\hline
\hline
Data series & \texttt{CTYPE1} & \texttt{CTYPE2} & \texttt{CRPIX1} & \texttt{CRPIX2} & \texttt{CRVAL1} & \texttt{CRVAL2} & \texttt{CDELT1} & \texttt{CDELT2} \\
\hline
Synoptic & \multirow{2}{*}{CRLN-CEA} & \multirow{2}{*}{CRLT-CEA} & \multirow{2}{*}{1800} & \multirow{2}{*}{720.5} & $360N-180$ & \multirow{2}{*}{0} & \multirow{2}{*}{-0.1} & \multirow{2}{*}{0.001389} \\
Synchronic & & & & & $360N-\phi_C-119.9$ & & & \\
\hline
\end{tabular}
\end{center}
\end{table}

%%%%%%%%%%%

\begin{itemize}

\item The \texttt{CTYPEn} keywords specify the coordinate system of the synoptic data (CEA). The prefixes \texttt{CRLN} and \texttt{CRLT} stand for Carrington longitude and latitude, respectively.

\item The \texttt{CRPIXn} keywords specify the reference pixel, with the lower left pixel at $(1,1)$. Note that the reference pixel in $x$-direction falls on a integer pixel, while in $y$-direction it falls between two pixels at the equator.

\item The \texttt{CRVALn} keywords specify the coordinate of the reference pixel. For $x$-axis, we use the ``Carrington time'', which is defined as $\Phi=360N-\phi$ for Carrington longitude $\phi$ and CR number $N$. Each column is marked by a unique $\Phi$, which increases from right to left on a map, and continues onto the next CR like real time. For $y$-axis, the center of the map is located at the equator, thus has a sine-latitude value 0.

In full-CR synoptic maps, the 1800th column corresponds to Carrington longitude $180\mdeg$, so its Carrington time is just $360N-180$. In synchronic maps, the central meridian (Stonyhurst longitude 0; Carrington longitude $\phi_C$) of the updated portion is placed at the 601st column. The Carrington longitude of the 1800th column is thus $119.9+\phi_C$, and the corresponding Carrington time is $360N-\phi_C-119.9$.

\item The \texttt{CDELTn} keywords specify the pixel sizes. In $x$-direction, the resolution is $0.1\mdeg$. The minus sign indicates that the Carrington time decreases from left to right. In $y$-direction, the resolution is $2.0/1440$ in sine-latitude.

\end{itemize}

%%%%%%%%%%%%%%%%%%%%%%%%%%%%%%%%%%%%%%%%%%%%%%%
%%%%%%%%%%%%%%%%%%%%%%%%%%%%%%%%%%%%%%%%%%%%%%%

\subsection{Correction Scheme}

We devised and applied a spatial-temporal interpolation correction scheme to fill in the polar fields that are missing or of poor quality for MDI \citep{sun2011}. The method is applied to HMI and is summarized as follows.

\begin{figure}[t!]
\centerline{\includegraphics[scale=0.9]{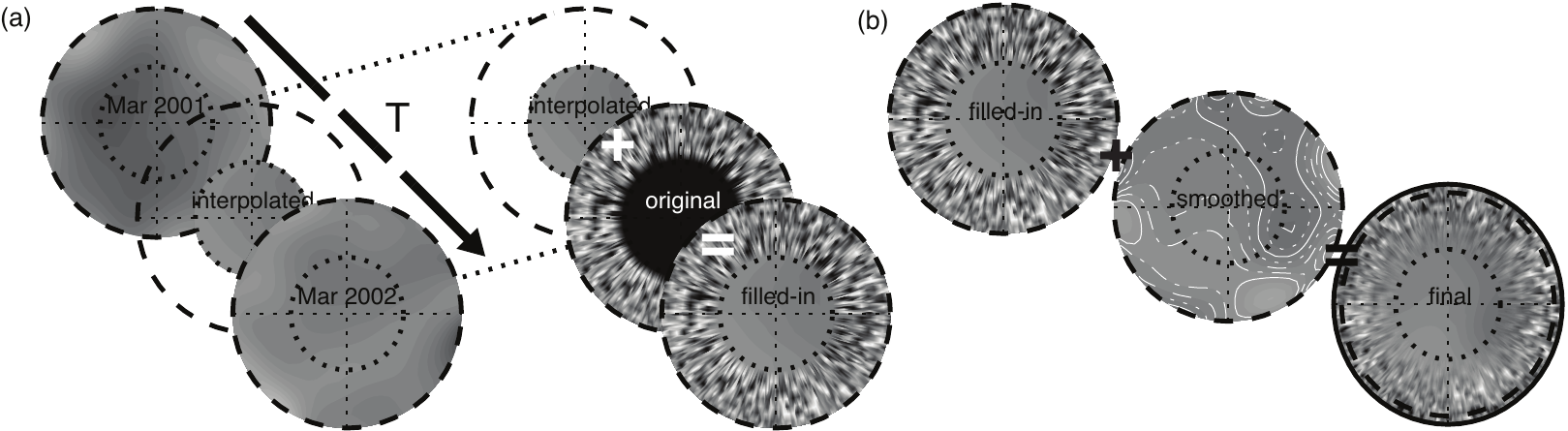}}
\vspace*{3mm}
\caption{\textbf{The polar field correction scheme.} The circular plates are stereoscopic views of the south poles above $62\mdeg$ latitude. Smaller dotted circles indicate $75\mdeg$. Black indicates missing data. (a) Processes of estimating the poorly measured pole-most pixels at a certain instant between March 2001 and March 2002 above $75\mdeg$. The temporally interpolated values are used to replace the pole-most pixels in observation. (b) Process of smoothly merging the filled-in data back to the map. The observed
field with gap filled is smoothed and merged back to reduce the noise level. Adapted from \citet{sun2011}.\label{f:corr}}
\end{figure}

We first create a ``polar field database'' using one set of best observed data each year, when the pole is tilted the most toward the Earth (Figure~\ref{f:corr}(a)). For north and south, we use the fiducial time September-06 02:00 UTC and March-08 02:00 UTC, respectively. Polar regions (above $60\mdeg$ latitude) centered at these times are extracted from the full-CR synoptic map series, and are rotated to have Carrington longitude $180\mdeg$ placed at the 1800th column. The data are subsequently remapped to a polar view using stereoscopic projection. The projected map is fitted with a 5th-order Chebyshev polynomial and thus effectively smoothed. The smoothed map is finally remapped back to the CEA coordinate.

To perform the correction, each synoptic map is designated with a time $T$ corresponding to \texttt{CRVAL1}, and each synchronic frame is designated with the time $T=T_C$ (the time of the new observation used to update the map). For each pole, we search for the polar field database for the two records $T_0$ and $T_1$ such that $T_0 \le T \le T_1$. For each pixel above $75\mdeg$ latitude in the original, we replace its value by temporally and linearly interpolating the two values from the two database records $T_0$ and $T_1$ at the same Carrington longitude/latitude (Figure~\ref{f:corr}(a)). To ease the discontinuity at $75\mdeg$, we smooth the map above $60\mdeg$ latitude (with values above $75\mdeg$ just corrected) using the Chebychev fitting method described in the previous step. To construct the final map, values above $75\mdeg$ are taken from this smoothed map. Values between $62\mdeg$ and $75\mdeg$ consist of a linear combination of the smoothed and the original map, transitioning from the former at higher latitudes to the latter at lower latitudes (Figure~\ref{f:corr}(b)). Because there can be more data missing in the synchronic frames, we use a lower filling-in latitude limit $72.5\mdeg$ (compared to $75\mdeg$ for synoptic maps).

For the earliest maps before the first record in the polar field database, the default interpolation method does not work. We resolve this by adding two well-observed polar field data points to the database using MDI synoptic maps from September 2009 and March 2010. These two maps are scaled by 0.82 after comparing the mean polar fields from MDI and HMI during CR 2097 - 2104 (Section~\ref{subsec:extrap}). The default interpolation method can now be used.

For the newest maps (all maps since the previous September for north and since the previous March for south), we generate the following ``provisional'' maps -- we create a predicted database record for the next year by scaling last record by factor of $k=1.21\overline{B}_{62}/\overline{B}_{75}$, where $\overline{B}_{62}$ is the mean field above $62\mdeg$, etc. This empirical approach mimics the effect of flux transport from the lower latitudes (Section~\ref{subsec:extrap}). With the observed and the predicted database records bracketing $T$, the default interpolation method can now be used. When the polar field database is updated (e.g. in September for north pole), these provisional maps are replaced by the ``definitive'' version.

%%%%%%%%%%%

%%%%%%%%%%%

%%%%%%%%%%%%%%%%%%%%%%%%%%%%%%%%%%%%%%%%%%%%%%%
%%%%%%%%%%%%%%%%%%%%%%%%%%%%%%%%%%%%%%%%%%%%%%%

\subsection{HMI Polar-Filled Synoptic Data Series}

We have created two new data series. One for full-CR synoptic maps (\href{http://jsoc.stanford.edu/ajax/lookdata.html?ds=hmi.synoptic_mr_polfil_720s}{\texttt{hmi.synoptic\_mr\_polfil\_720s}}), and one for daily-updated synchronic frames (\href{http://jsoc.stanford.edu/ajax/lookdata.html?ds=hmi.mrdailysynframe_polfil_720s}{\texttt{hmi.mrdailysynframe\_polfil\_720s}}).

The polar field corrected synoptic maps are generated along with the original maps (\href{http://jsoc.stanford.edu/ajax/lookdata.html?ds=hmi.synoptic_mr_720s}{\texttt{hmi.synoptic\_mr\_720s}}). The newest map are provisional, and are specified by the non-zero keywords \texttt{N\_EXTRAP} and \texttt{S\_EXTRAP} for north and south pole, respectively. After September each year, the north polar field database are updated. All maps since the previous September are regenerated, and \texttt{N\_EXTRAP} is set to 0. The same procedures are performed for the south every March. Maps more than one year old are definitive. Because we do not use  \texttt{N\_EXTRAP} and \texttt{S\_EXTRAP} as identifier, a simple query in the SDO database will return only the most recently updated version. We expect the differences to be small (Section~\ref{subsec:extrap}), unless the large-scale polar field changes rapidly.
 
The newest synchronic frames will always have the provisional polar field correction. Nevertheless, because the data series was first processed in December 2017, north pole prior to September 2017 and south pole prior to March 2017 will have the definitive correction at that time.

%%%%%%%%%%%%%%%%%%%%%%%%%%%%%%%%%%%%%%%%%%%%%%%%%%%%%%%%%%%%%%
%%%%%%%%%%%%%%%%%%%%%%%%%%%%%%%%%%%%%%%%%%%%%%%%%%%%%%%%%%%%%%

\section{Results}
\label{sec:result}

%%%%%%%%%%%%%%%%%%%%%%%%%%%%%%%%%%%%%%%%%%%%%%%
%%%%%%%%%%%%%%%%%%%%%%%%%%%%%%%%%%%%%%%%%%%%%%%

\subsection{Mean Polar Field}
\label{subsec:mf}

\begin{figure}[t!]
\centerline{\includegraphics[scale=0.9]{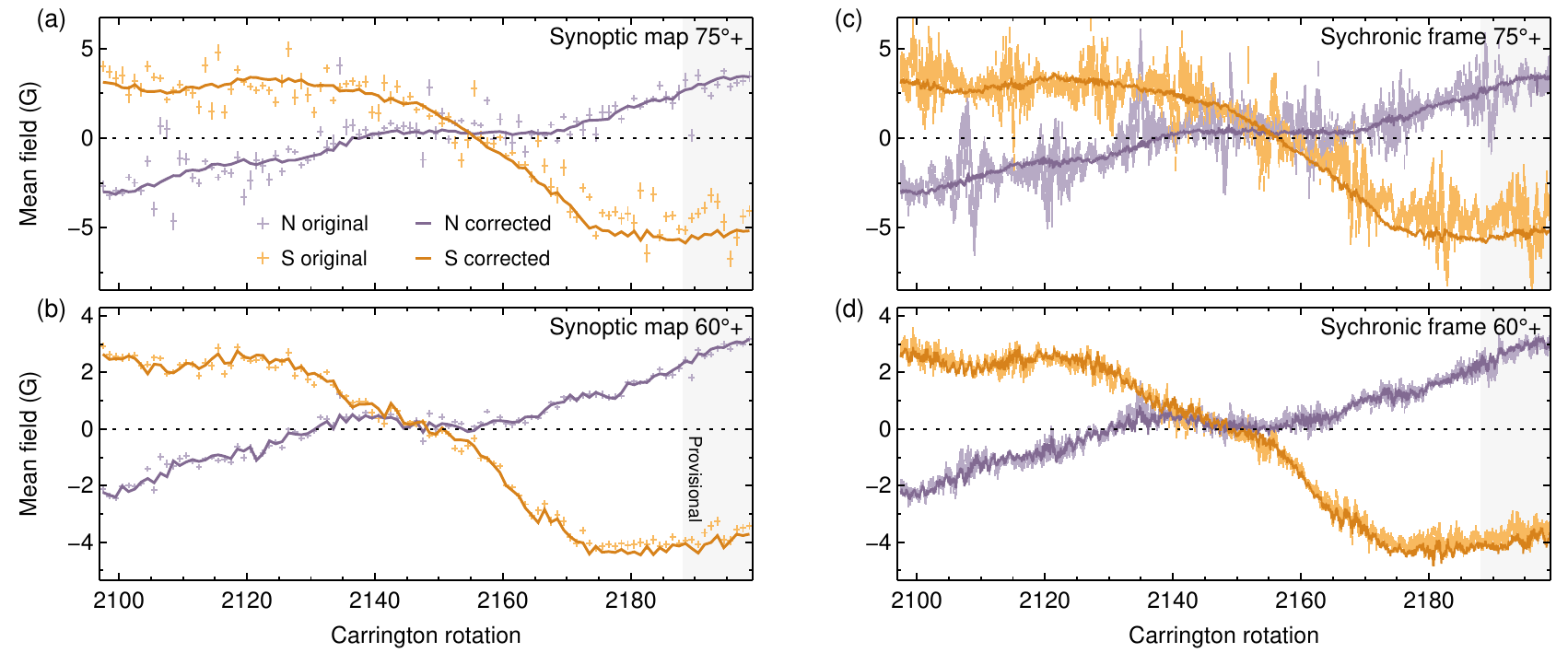}}
\caption{\textbf{Mean polar field before and after correction.} (a) Mean polar field above 75$\mdeg$ latitude in north (purple) and south (orange) for original (crosses) and corrected (curve) synoptic maps from CR 2097 to 2198. The vertical error bars show $5\sigma$ standard error of the mean. The shaded region denotes the period after September 2017 when the correction is ``provisional'' and will be updated as new observations become available. (b) Similar to (a), for polar field above 60$\mdeg$ latitude. (c) and (d), similar to (a) and (b), but for daily-updated synchronic frames from 2010-06-01 to 2017-12-28. \label{f:result}}
\end{figure}

We apply the aforementioned correction scheme to HMI synoptic maps and daily updated synchronic frames. The mean polar field in the corrected maps matches the raw maps closely (Figure~\ref{f:result}) and agree with the values calculated directly from magnetograms \citep{sun2015}. The difference is greater at higher latitudes (above 75$\mdeg$) compared lower latitudes (above 60$\mdeg$) as expected. For synoptic maps, the root-mean-square (rms) difference is 0.73 G and 0.98 G for north ($\delta B_{N75}$) and south ($\delta B_{S75}$) above 75$\mdeg$, respectively. For synchronic frames, the rms is 0.70 G and 0.99 G for $\delta B_{N75}$ and $\delta B_{S75}$, respectively. The scatter increases significantly when the pole starts to tilt away from the Earth. For example, the rms $\delta B_{N75}$ for $b \le -5\mdeg$ ($b$ is tilt of solar rotational axis, positive when the north is tilted toward the Earth) is 1.34 G for synoptic maps, about twice the overall rms.

As of December 2017 when the calculation was initially performed, the north pole was updated through September 2017, and the south pole updated through March 2017. Therefore maps after these dates had provisional corrections (shaded bands in Figure~\ref{f:result}). The corrected southern polar fields appear to be slightly but consistently stronger than raw measurements, which may be caused by an overestimate of the southern polar field next March in our scheme (Section~\ref{subsec:extrap}).

%%%%%%%%%%%

%%%%%%%%%%%

\begin{figure}[t!]
\centerline{\includegraphics[scale=0.9]{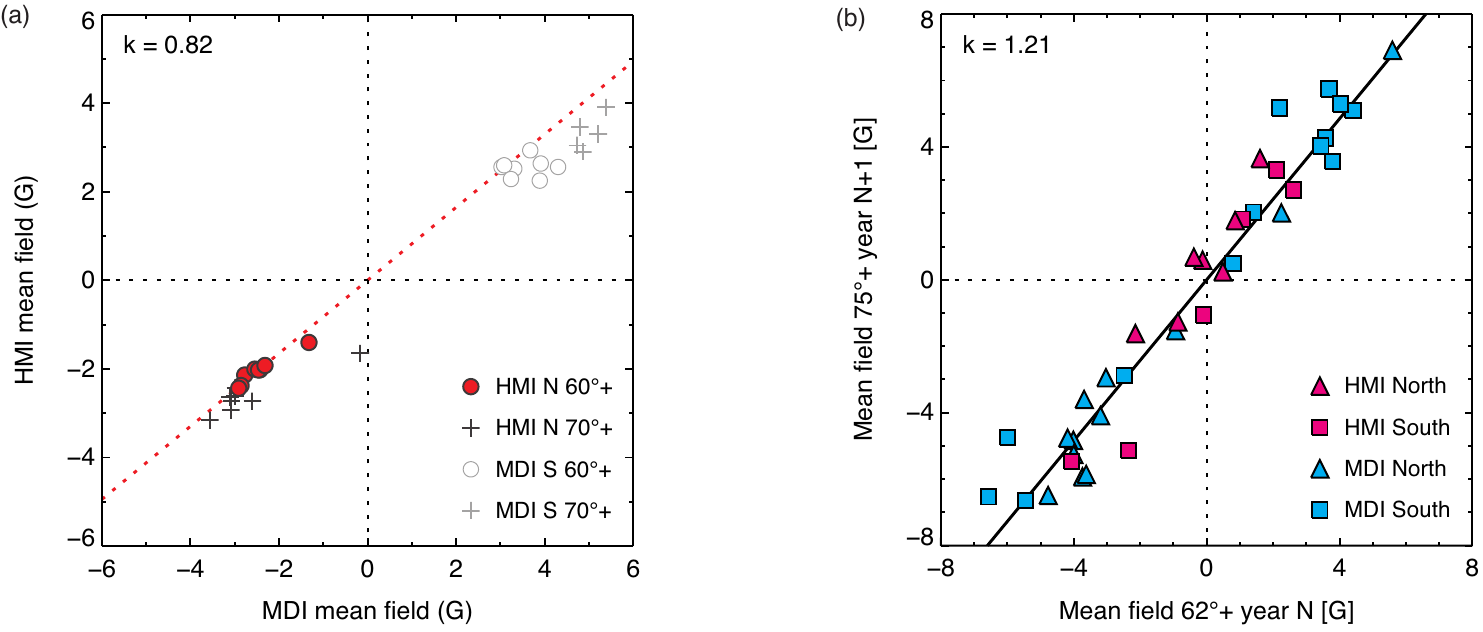}}
\caption{\textbf{Extrapolation scheme.} (a) Comparison of mean polar field from MDI and HMI from 2010 May to November (CR 2097 - 2104). Only the northern mean field above 60$\mdeg$ latitude is used to determine the scaling due to possible bias in MDI at high latitudes when the pole is tilted away from the Earth. (b) Predicting the polar field above 75$\mdeg$ latitude of year $N+1$ using observed values above 60$\mdeg$ of year $N$.\label{f:extrap}}
\end{figure}

%%%%%%%%%%%%%%%%%%%%%%%%%%%%%%%%%%%%%%%%%%%%%%%
%%%%%%%%%%%%%%%%%%%%%%%%%%%%%%%%%%%%%%%%%%%%%%%

\subsection{Extrapolation}
\label{subsec:extrap}

We use MDI measurements (September 2009 for north and March 2010 for south) to constrain the polar field before the first HMI data (September 2010 for north and March 2011 for south). Because MDI and HMI data have a non-unity scaling that varies slightly in space \citep{liuy2012}, we compare the mean polar fields from these two instruments from May to November 2010 (Figure~\ref{f:extrap}(a)). As the south pole was starting to tilt away from the Earth and thus suffers from larger uncertainly at higher latitudes, we use the mean field above 60$\mdeg$ in the north alone for comparison, from which we get a best-fit slope of 0.82.

\begin{figure}[t!]
\centerline{\includegraphics[scale=0.9]{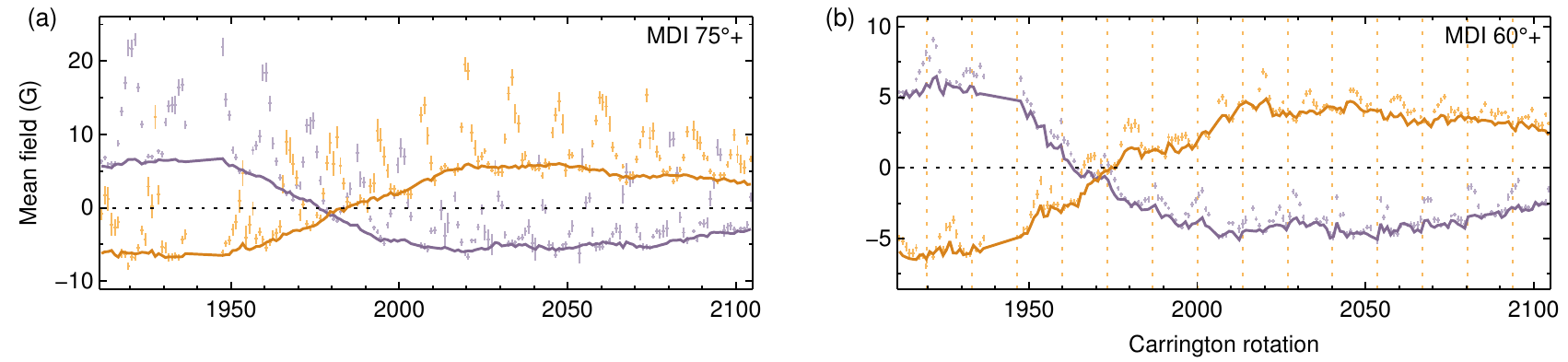}}
\caption{\textbf{Mean polar field before and after correction for MDI.} (a) (b) Similar to Figure~\ref{f:result}(a) (b), but for MDI. The fiducial southern observations in March each year is denoted by vertical dotted lines in (b). \label{f:mdi}}
\end{figure}

It is obvious that the MDI southern polar fields is stronger than the scaling we adopt (Figure~\ref{f:extrap}(a), first quadrant). We tentatively consider this a systematic bias for the following reason. We find that he MDI mean polar fields almost always have a positive offset compared to the corrected version, and the offset increases when the pole becomes more tilted away from the Earth (Figure~\ref{f:mdi}). This is in contrast with HMI (Figure~\ref{f:result}), where the offset can be either positive or negative. Such a bias is unlikely to be physical. Investigation of its origin is underway.

%%%%%%%%%%%

%%%%%%%%%%%

We verify the forward-extrapolation scheme from \citep{sun2011} using both MDI and HMI data (Figure ~\ref{f:extrap}(b)). The mean field above 62$\mdeg$ from year $N$, $B_{62}(N)$, is well correlated with the mean field above 75$\mdeg$ from year $N+1$, $B_{75}(N+1)$, with a best-fit slope of 1.21. We thus use the value $1.21B_{62}(N)$ as a predictor of $B_{75}(N+1)$. The rms difference, as an estimate of error, is 1.00 G. Values that deviate more than 2 times the rms occurred in the south for the year 2000, 2004, and 2016, when the polar field evolution was relatively fast.

%%% %%%%%%%%%%%%%%%%%%%%%%%%%%%%%%%%%%%%%%%%%%%%%%%%%%%%%%%%%%%
%% Bibliography
%
% Using BibTeX
%
\bibliographystyle{apj} 
\bibliography{polarfill}  

\begin{thebibliography}{}
\expandafter\ifx\csname natexlab\endcsname\relax\def\natexlab#1{#1}\fi

\bibitem[{{Liu} {et~al.}(2012){Liu}, {Hoeksema}, {Scherrer}, {Schou},
  {Couvidat}, {Bush}, {Duvall}, {Hayashi}, {Sun}, \& {Zhao}}]{liuy2012}
{Liu}, Y., {Hoeksema}, J.~T., {Scherrer}, P.~H., {et~al.} 2012, Sol. Phys.,
  279, 295

\bibitem[{{Sun} {et~al.}(2015){Sun}, {Hoeksema}, {Liu}, \& {Zhao}}]{sun2015}
{Sun}, X., {Hoeksema}, J.~T., {Liu}, Y., \& {Zhao}, J. 2015, Astronphys. J.,
  798, 114

\bibitem[{{Sun} {et~al.}(2011){Sun}, {Liu}, {Hoeksema}, {Hayashi}, \&
  {Zhao}}]{sun2011}
{Sun}, X., {Liu}, Y., {Hoeksema}, J.~T., {Hayashi}, K., \& {Zhao}, X. 2011,
  Sol. Phys., 270, 9

\bibitem[{{Thompson}(2006)}]{thompson2006}
{Thompson}, W.~T. 2006, Astron. Astrophys., 449, 791

\end{thebibliography}

\end{document}